\pdfoutput=1

\documentclass{appolb}

\usepackage{epsfig}
\usepackage{hyperref}

\begin{document}

\title{Wounded quarks at the LHC%
\thanks{Presented by WB at Critical Point and the Onset of Deconfinement (CPOD 2016), Wroc\l{}aw, Poland, 30 May - 4 June 2016.}}

\author{Wojciech Broniowski$^{1,2}$,  Piotr Bo\.zek$^3$, Maciej Rybczy\'nski$^1$
\address{$^1$ Institute of Physics, Jan Kochanowski University, 25-406 Kielce, Poland \\
               $^2$ The H. Niewodnicza\'nski Institute of Nuclear Physics, Polish Academy of Sciences, 31-342 Cracow, Poland \\
               $^3$ AGH University of Science and Technology, Faculty of Physics and Applied Computer Science, 30-059 Cracow, Poland }
}
\maketitle

\begin{abstract}
We review the results of the wounded quark model, with a stress on 
eccentricity observables in small systems. A new element is a presentation of symmetric cumulants
for the elliptic and triangular flow correlations, obtained in the wounded-quark approach.
\end{abstract}


\bigskip \bigskip
  
This talk is largely based on~\cite{Bozek:2016kpf} where more details and results can be found.  
Historically, the wounded quark model~\cite{Bialas:1977en,Bialas:1977xp,Bialas:1978ze,Anisovich:1977av} was introduced shortly after the 
phenomenological success of wounded nucleon model~\cite{Bialas:1976ed}. The approach stems from the Glauber model~\cite{Glauber:1959aa} adapted to inelastic 
production~\cite{Czyz:1969jg}. Our results contribute to the on-going discussion on the nature of the initial stages of the ultra-relativistic collision and the 
relevant degrees of freedom taking part in the early production of entropy/energy in the fireball: Are these nucleons, quarks, partons, random fields, 
gluonic hot-spots? As the combinatorics of the production depends on the number of constituents, showing in the dependence of the particle production 
on centrality, it offers a possibility to assess the number of active constituents without a detailed reference to their physical nature.

In~\cite{Eremin:2003qn} it was noticed that the RHIC data can be explained within a wounded quark model, where 
the linear scaling $\frac{dN_{\rm ch}}{d\eta} \sim Q_{\rm W}$ is used, with $Q_W$ denoting the number of wounded quarks. The model was further 
advocated by the PHENIX Collaboration~\cite{Adler:2013aqf,Adare:2015bua}.
The wounded-quark scaling also works for the SPS~\cite{KumarNetrakanti:2004ym}. 
More recent development in this direction can be found in~\cite{Loizides:2014vua,Adler:2013aqf,Adare:2015bua,Lacey:2016hqy,Zheng:2016nxx,Mitchell:2016jio,Loizides:2016djv}.

\begin{figure}
\begin{center}
\includegraphics[width=0.7 \textwidth]{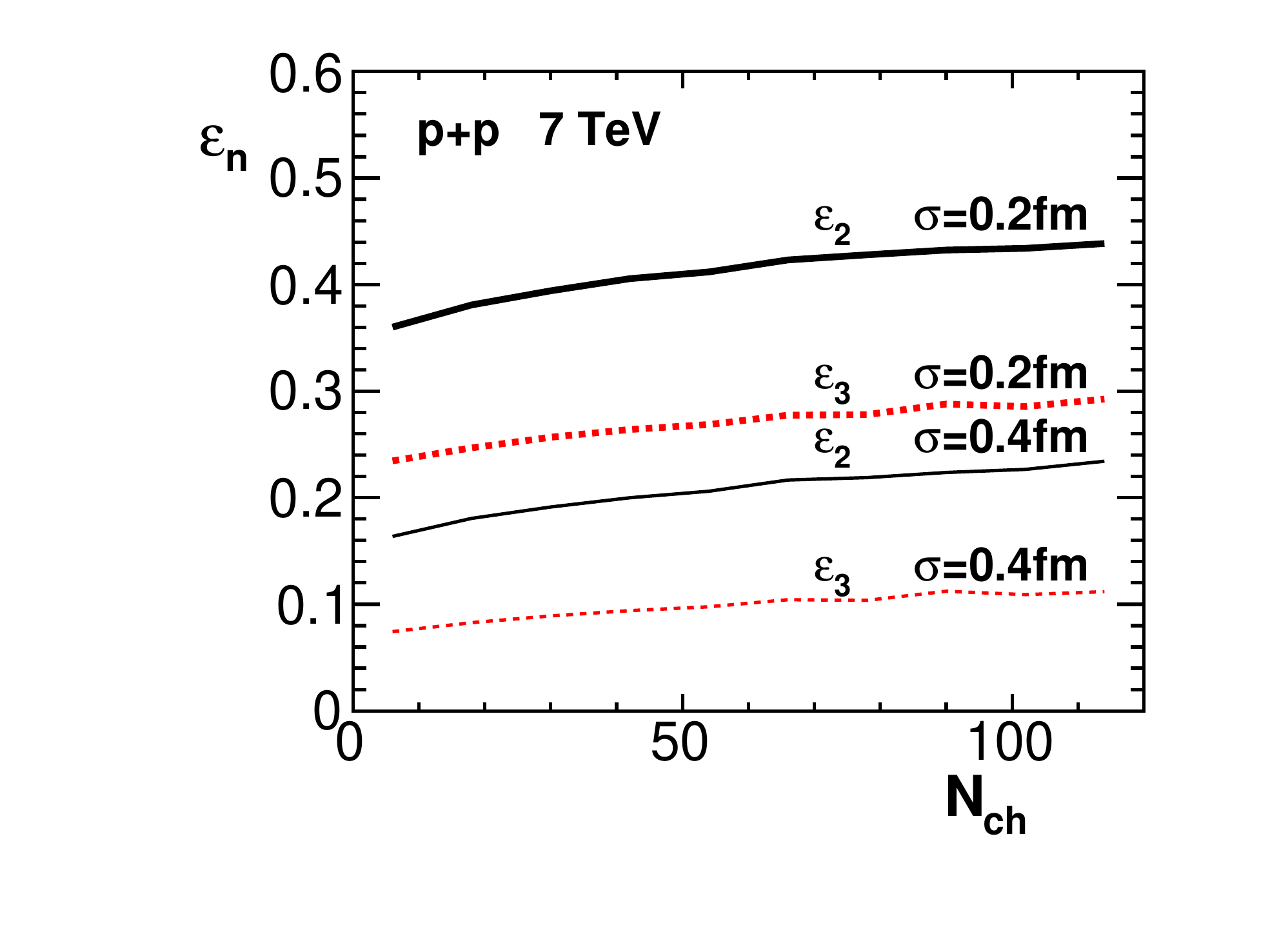} 
\end{center}
\vspace{-11mm}
\caption{Ellipticity $\varepsilon_2$ (solid lines) and triangularity $\varepsilon_3$ (dotted lines) 
of the fireball in p+p collisions at $\sqrt{s}=7$~TeV. The thin and thick lines correspond to the Gaussian smearing parameter $\sigma=0.4$~fm and $0.2$~fm,
respectively. The eccentricities are plotted as functions of the mean charged multiplicity  at $|\eta| < 2.4$. 
\label{fig:pp}}
\end{figure}

Modeling at the subnucleonic level allows for examination of p+p collisions with the techniques typically used for larger systems. In Fig.~\ref{fig:pp} we show
eccentricities resulting from the wounded quarks in fireballs formed in p+p collisions at the LHC. We note that both the ellipticity and triangularity are large, hence may
lead to substantial harmonic flow, in accordance to the collectivity mechanism expected for collisions 
of small systems~\cite{Bozek:2011if,Bozek:2012gr,Bozek:2013uha,Bozek:2013ska} with sufficiently high multiplicity. The two sets of curves, thinner and thicker, correspond 
to two values of the source smearing parameter~\cite{Bozek:2016kpf}. We note that the eccentricities do not strongly depend on the number of produced particles (centrality), 
which shows that they originate from fluctuations and not the geometry of the collision.

As a new result of the simulations in the wounded-quark model as implemented in~\cite{Bozek:2016kpf}, we present the 
ellipticity-triangularity correlations. We use the measure in the form of a symmetric cumulant, as introduced by the ALICE 
Collaboration~\cite{ALICE:2016kpq}:
\begin{eqnarray}
\frac{SC(a,b)}{\langle a^2\rangle \langle b^2\rangle }=\frac{\langle a^2 b^2\rangle -\langle a^2\rangle \langle b^2\rangle }{\langle a^2\rangle \langle b^2\rangle }.
\end{eqnarray}
As for a given reaction the elliptic and triangular flow coefficients are roughly proportional to the corresponding eccentricities, 
\begin{eqnarray}
v_n \simeq \kappa(n,c) \epsilon_n, \;\; n=2,3, 
\end{eqnarray}
with $c$ indicating centrality, one obtains the approximate relation
\begin{eqnarray}
\frac{SC(v_2,v_3)}{\langle v_2^2\rangle \langle v_3^2\rangle }\simeq\frac{SC(\epsilon_2,\epsilon_3)}{\langle \epsilon_2^2\rangle \langle \epsilon_3^2\rangle },
\end{eqnarray}
where the dependence on $\kappa(n,c)$ has canceled out.

\begin{figure}
\begin{center}
\includegraphics[width=.67\textwidth]{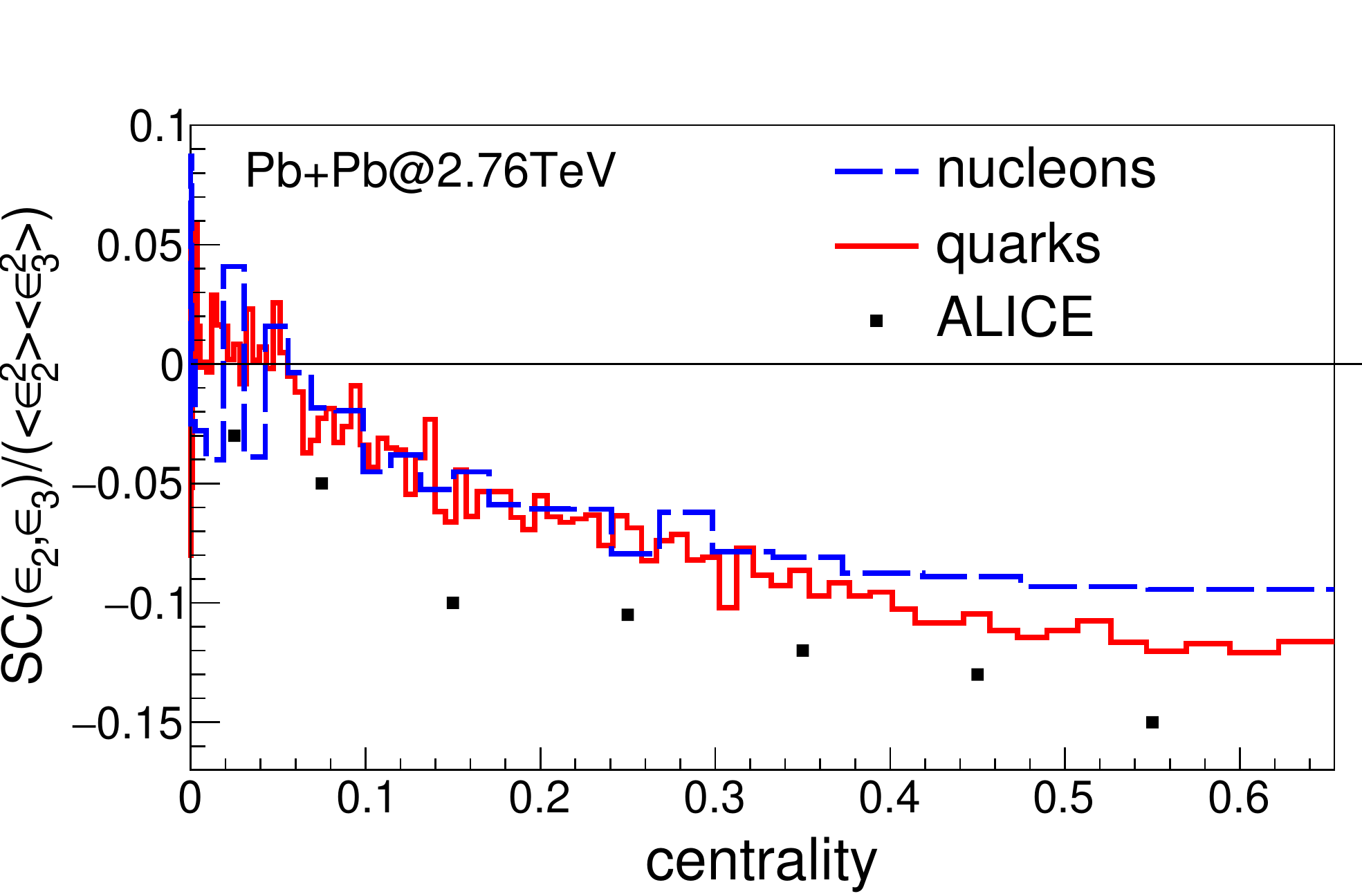}
\end{center}
\caption{The correlation between the elliptic and triangular eccentricities for Pb+Pb collisions at the LHC, expressed via the symmetric cumulant. 
The lines indicate the calculations in the wounded nucleon and 
wounded quark models, whereas the points correspond to the ALICE data~\cite{ALICE:2016kpq}. \label{fig:alice}}
\end{figure}

\begin{figure}
\begin{center}
\includegraphics[width=.67 \textwidth]{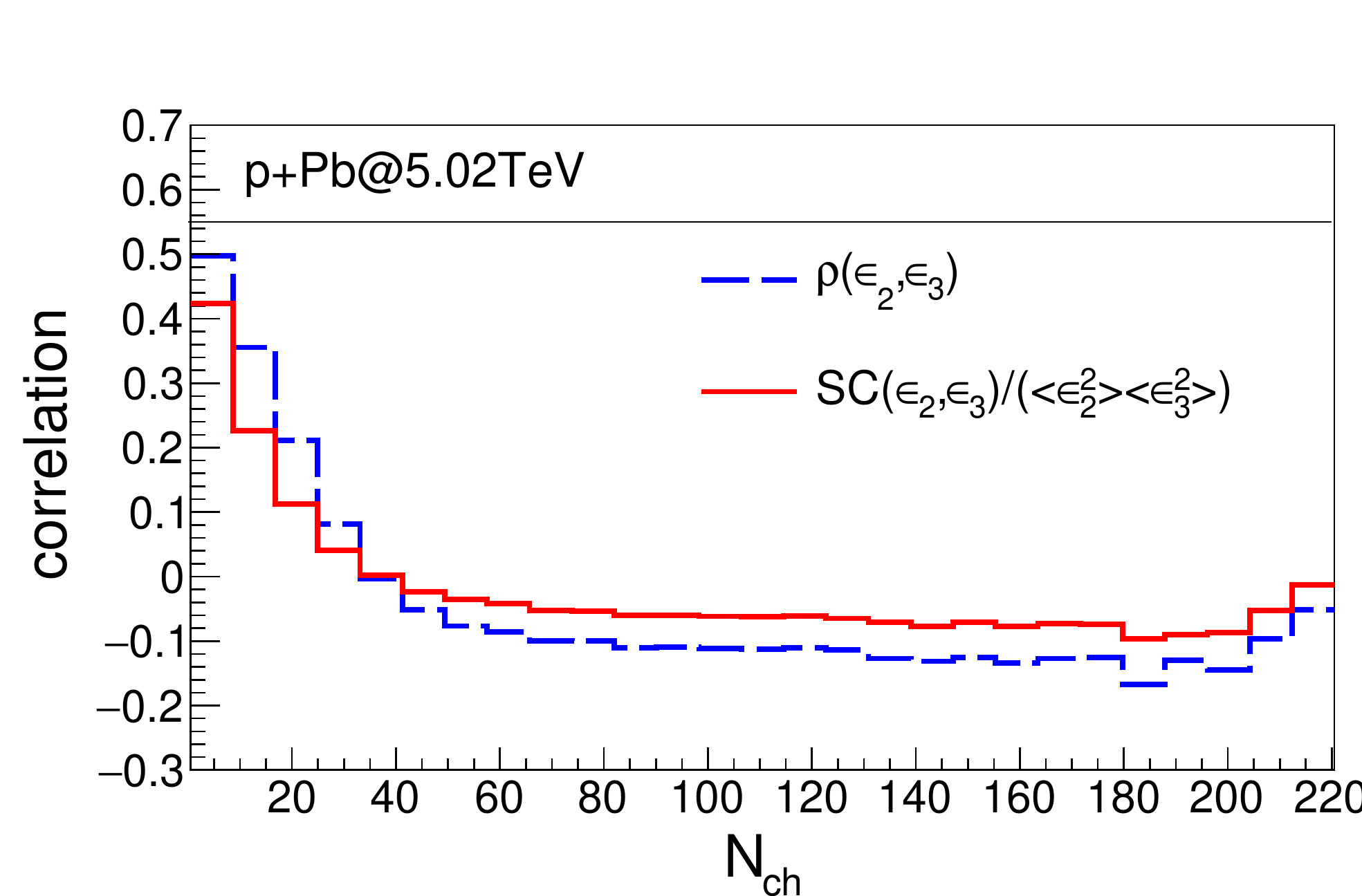} \\ \vspace{-12mm} \includegraphics[width=.67 \textwidth]{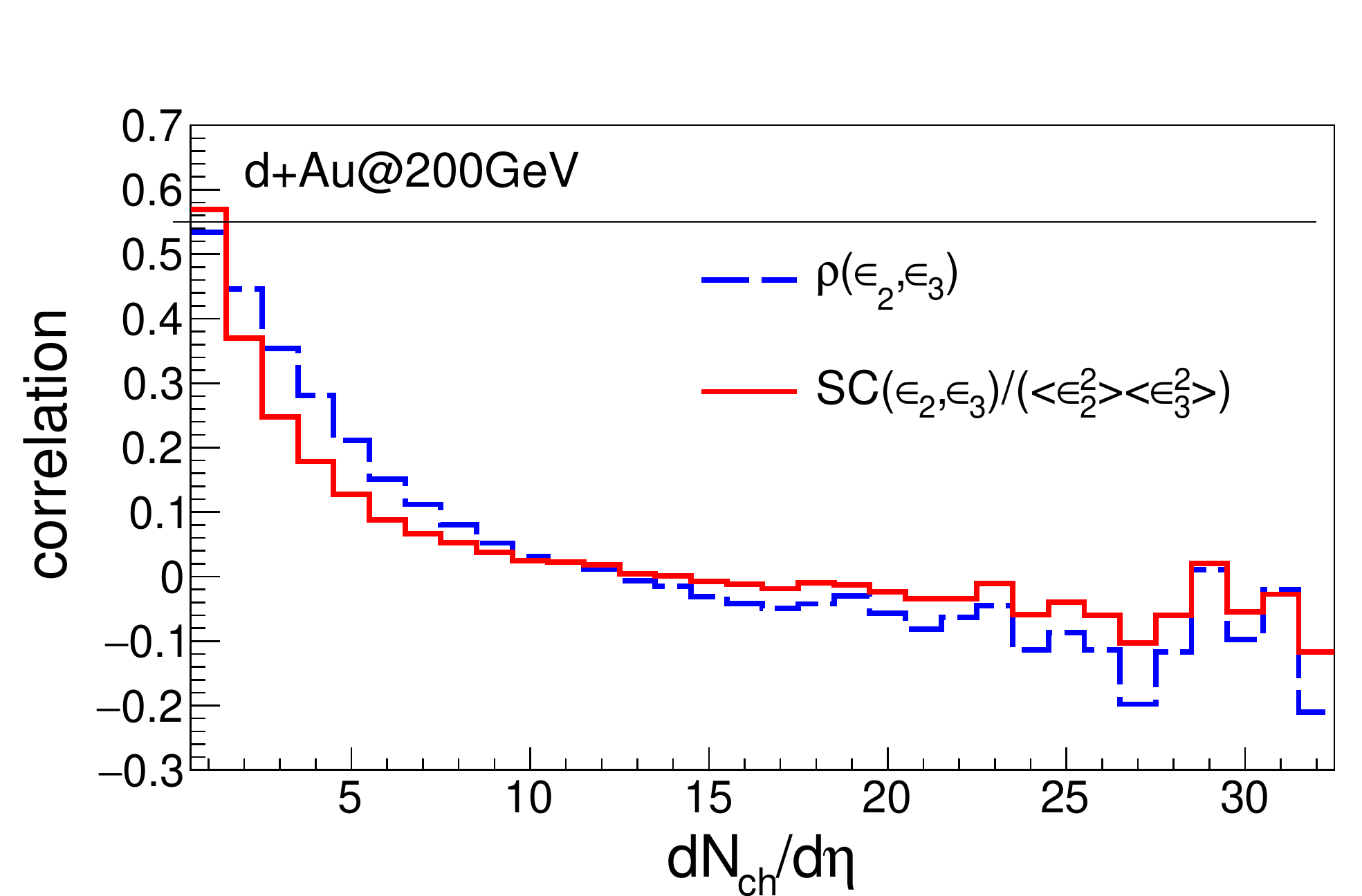} \\ \vspace{-12mm} \includegraphics[width=.67 \textwidth]{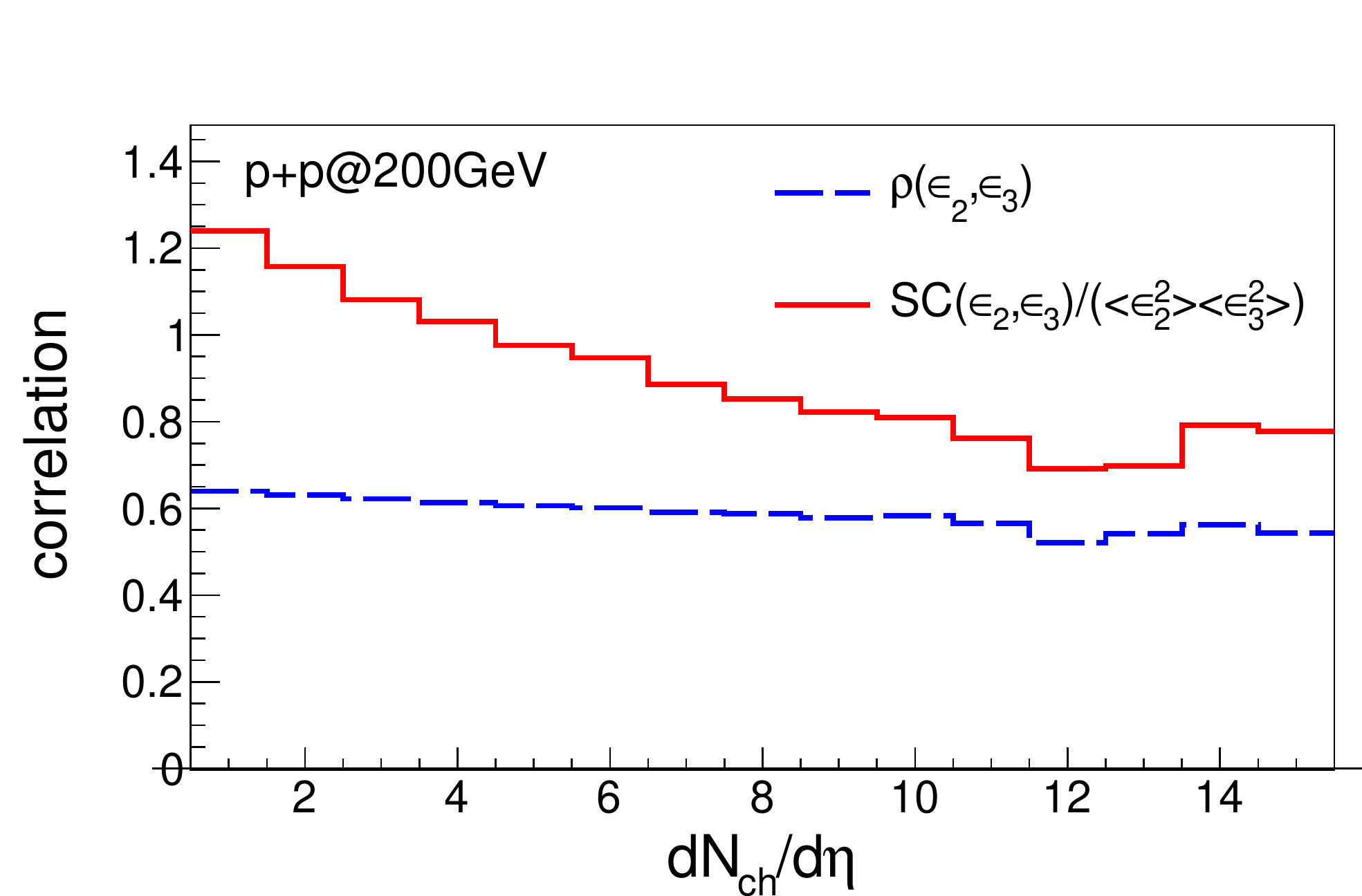}
\end{center}
\caption{Predictions for the correlation measures between the elliptic and triangular eccentricities in small systems, 
obtained from the wounded quark model for three sample reactions. \label{fig:sc}}
\end{figure}

The results for Pb+Pb collisions at the LHC are presented in Fig.~\ref{fig:alice}. We note that the predictions from the wounded-nucleon and wounded-quark models
are similar to each other and follow the trend of the data~\cite{ALICE:2016kpq}, indicated with points. 
The negative sign and the fall-off with centrality are properly reproduced.

The predictions for small systems are shown in Fig.~\ref{fig:sc}.
We also plot there the standard Pearson's correlation coefficient 
\begin{eqnarray}
\rho(a,b)=\frac{\langle a b\rangle  -\langle a\rangle \langle b\rangle }{\sqrt{(\langle a^2\rangle -\langle a\rangle ^2)(\langle b^2\rangle -\langle b\rangle ^2)}}.
\end{eqnarray}
For p+Pb and d+Au reactions the coefficient $\rho(\epsilon_2,\epsilon_3)$  follows closely the symmetric cumulant measure, whereas for p+p collision it is substantially different. Such 
correlations as displayed in Fig.~\ref{fig:sc}, which are large in the considered approach, may be checked in future data analyses of flow correlations in small systems.

We wish to thank Ante Bilandzi\'c for a useful discussion concerning symmetric cumulants. This eesearch was supported by the Polish National Science Centre grants 
2015/17/B/ST2/00101, 2012/06/A/ST2/00390, and \mbox{2015/18/M/ST2/00125}.

\bibliography{hydr}

\end{document}